\documentclass[aps,twocolumn,nofootinbib]{revtex4}
\usepackage{graphicx}
\usepackage{adjustbox}
\usepackage{dcolumn}
\usepackage{bm}
\usepackage{amsmath}
\usepackage{bbold}
\usepackage{multirow}
\usepackage{amsmath}
\usepackage[letterpaper, hmargin = {2cm}, vmargin = {2cm}]{geometry}
\usepackage{tablefootnote}
\usepackage[ulem=normalem]{changes}
\usepackage{float}

\newcommand{\Gj}[6]{ \begin{Bmatrix}
  #1 & #2 & #3 \\
  #4 & #5 & #6
 \end{Bmatrix}}

\begin{document}

\title{The core of $^{25}$F studied by the $^{25}$F$(-p)$ proton-removal reaction}

\author{H.~L.~Crawford$^{1}$\footnote{Corresponding author: hlcrawford@lbl.gov}}
\author{M.~D.~Jones$^{1,2}$}% Comments received; DONE \footnote{Present address:}}
\author{A.~O.~Macchiavelli$^{1,3}$}
\author{P.~Fallon$^{1}$}
\author{D.~Bazin$^{4}$}
\author{P.~C.~Bender$^{4}$\footnote{Present address: Department of Physics and Applied Physics, University of Massachusetts Lowell, Lowell, MA 01854, USA}}
\author{B.~A.~Brown$^{4,5}$} % Comments received; DONE
\author{C.~M.~Campbell$^{1}$} % Confirmed
\author{R.~M.~Clark$^{1}$}  % Comments received; DONE
\author{M.~Cromaz$^{1}$}
\author{B.~Elman$^{4,5}$}
\author{A.~Gade$^{4,5}$} % Comments received; DONE
\author{J.~D.~Holt$^{6}$}
\author{R.~V.~F.~Janssens$^{2}$} % Comments received; DONE
\author{I.~Y.~Lee$^{1}$} % Comments received; DONE
\author{B.~Longfellow$^{4,5}$\footnote{\label{note1}Present address: Lawrence Livermore National Laboratory, Livermore, CA 94550, USA}}
\author{S.~Paschalis$^{7}$} % Comments received; DONE
\author{M.~Petri$^{7}$} % Comments received; DONE
\author{A.~L.~Richard$^{4\ddag}$}
\author{M.~Salathe$^{1}$} % Comments received; DONE
\author{J.~A.~Tostevin$^{8}$} % Comments received; DONE
\author{D.~Weisshaar$^{4}$} % Confirmed
\affiliation{$^{1}$Nuclear Science Division, Lawrence Berkeley National Laboratory, Berkeley, CA 94720, USA}
\affiliation{$^{2}$Department of Physics and Astronomy, University of North Carolina at Chapel Hill, Chapel Hill, NC 27559-3255, USA and Triangle Universities Nuclear Laboratory, Duke University, Durham, NC 27708-0308, USA}
\affiliation{$^{3}$Physics Division, Oak Ridge National Laboratory, Oak Ridge, TN 37831, USA}
\affiliation{$^{4}$National Superconducting Cyclotron Laboratory, Michigan State University, East Lansing, MI 48824, USA}
\affiliation{$^{5}$Department of Physics and Astronomy, Michigan State University, East Lansing, MI 48824, USA}
\affiliation{$^{6}$TRIUMF, 4004 Wesbrook Mall, Vancouver, BC V6T 2A3, Canada}
\affiliation{$^{7}$Department of Physics, University of York, Heslington, York YO10 5DD, United Kingdom}
\affiliation{$^{8}$Department of Physics, University of Surrey, Guildford, Surrey GU2 7XH, United Kingdom}
\date{\today}

\begin{abstract}
The $^9$Be$(^{25}$F($5/2^+),^{24}$O)X proton-removal reaction was studied at the NSCL using the S800 spectrometer.  The experimental spectroscopic factor for the ground-state to ground-state transition indicates a substantial depletion of the proton $d_{5/2}$ strength compared to shell-model expectations, similar to the findings of an inverse-kinematics $(p,2p)$ measurement performed at RIBF. The $^{25}$F to $^{24}$O ground-states overlap is considerably less than anticipated if the core nucleons behaved as rigid, doubly-magic $^{24}$O within $^{25}$F. We interpret the new results within the framework of the Particle-Vibration Coupling (PVC) model, of a $d_{5/2}$ proton coupled to a quadrupole phonon of an effective core. This approach provides a good description of the experimental data, requiring an effective $^{24}$O* core with a phonon energy of $\hbar\omega_2$= 3.2 MeV and a $B(E2) \approx 2.7$ W.u. -- softer and more collective than a bare $^{24}$O. Both the Nilsson deformed mean field and the PVC models appear to capture the properties of the effective core of $^{25}$F, suggesting that the additional proton polarizes $^{24}$O in such a way that it becomes either slightly deformed or a quadrupole vibrator.
\end{abstract}

%\pacs{21.10.Jx, 21.60.Ev, 25.45.-z, 27.20.+n}

\maketitle

\section{Introduction}

The nature of shell closures and the persistence of magic numbers in exotic neutron-rich nuclei is a fundamental question of major importance in nuclear physics.  Studies aiming to identify and understand the evolution of shell structure and collectivity, moving away from $\beta$-stability, have attracted major efforts worldwide.

A unique opportunity to study these effects can be found in the oxygen isotopic chain, with a closed $Z$=8 proton shell.  Experimental work carried out at NSCL~\cite{Hoffman09} and GSI~\cite{Kanungo2009} revealed that $^{24}$O, located at the neutron dripline, is a doubly-magic nucleus, with $Z$=8 and $N$=16, confirming earlier experimental indications~\cite{Ozawa2000,Stanoiu2004} and theoretical predictions~\cite{Brown05}.  The structure of the neutron-rich oxygen isotopes has come to be understood in terms of interactions between valence neutrons and the core nucleons driving modifications to effective single-particle orbital energies, providing an important testing ground for shell-model interactions and \textit{ab-initio} descriptions of medium-mass nuclei.  These nuclei have provided an important benchmark to study the effects of three-nucleon (3N) forces in determining the location of the neutron dripline~\cite{Otsuka2010,Hebeler2015}.  One of the most dramatic manifestations of the unique role of the strong proton-neutron force appears when one compares the oxygen and the $Z$=9 fluorine isotopes.  With just one more proton than oxygen, the neutron dripline in F extends seven neutrons beyond $^{24}$O to $^{31}$F, as the strong overlap between the spin-orbit partners $\pi 1d_{5/2}$-$\nu 1d_{3/2}$ lowers (binds) the $\nu 1d_{3/2}$ orbital in fluorine and changes the shell gap.

\textit{Ab-initio} calculations with two-nucleon (NN) and 3N interactions, which were successfully used to describe the location of the oxygen dripline at $^{24}$O~\cite{Bogner2014}, are now being extended to the fluorine isotopic chain~\cite{Bogner2014, Stroberg2016}.  Given the sensitivity to the subtle interplay of nuclear forces in this region, it is important to obtain detailed structure data on neutron-rich O and F isotopes to test theories that aim to describe and predict the structure of nuclei out to the dripline.  One of the most sensitive tests is through direct reaction experiments, measuring exclusive cross sections to final states where a nucleon has been added or removed and their derived spectroscopic factors and associated level occupancies.

A recent RIKEN/RIBF measurement~\cite{Tang2020-PRL} explored this question via the spectroscopic factors connecting the $^{25}$F ground state and $^{24}$O final states. Using the inverse kinematics $^{25}$F$(p,2p)$ reaction at 270 MeV/nucleon on a liquid hydrogen target, Tang \textit{et al.} measured partial cross sections to the $^{24}$O ground state, the only particle-bound final state, and to unbound excited states. They reported that, taken together, these cross sections accounted for the total expected $1d_{5/2}$ proton single-particle strength, but that the $^{25}$F ground state differs significantly from a dominantly $^{24}$O$_{gs} + p$ configuration.  Specifically, the experimental $^{25}$F ground state to the $^{24}$O ground state spectroscopic factor C$^{2}$S$_{exp}$, based on a computed theoretical $(p,2p)$ cross section into their very restricted detection geometry, was 0.36(13). The interpretation was therefore that the core nucleons in $^{25}$F have only a $\sim$36\% probability of being found in the $^{24}$O ground state and that the core nucleon configurations are dominated by excited states. This suggests that the single, additional 1$d_{5/2}$ proton in $^{25}$F substantially alters the structure of the core nucleons. Such a core polarization was not predicted by large-scale shell model calculations employing state-of-the-art phenomenological interactions in this mass region~\cite{Tang2020-PRL}.

Given this significant departure from expectations, further experimental study of the cross section and spectroscopic factor for proton removal from $^{25}$F is warranted. We employ a different and complementary direct reaction in which the proton is removed in fast collisions of an intermediate-energy $^{25}$F secondary beam with a target of light nuclei, here $^{9}$Be. We refer to this reaction mechanism as proton-removal, reserving the term knockout for the quasifree $(p,2p)$ process. The present measurement adds substantial statistics to an earlier proton-removal experiment, performed at NSCL at a lower beam energy of 50 MeV/nucleon on a carbon target~\cite{Thoennessen2003}. Throughout this work, our analysis uses the sudden, eikonal removal-reaction model of Ref. \cite{Tostevin2021-PRC}. When this earlier experiment is so-analysed, which predicts a single-particle removal cross section of 13.3~mb, one obtains an experimental spectroscopic factor of 0.29(5). This, and our new result discussed in Section III, in common with Tang {\em et al.} \cite{Tang2020-PRL}, indicate strong suppression of the proton $d_{5/2}$ strength. We note that the secondary beam energy in the earlier NSCL measurement \cite{Thoennessen2003} is at the lower end of values for which the eikonal, dynamical model used is considered reliable in cases where the removed nucleon is reasonably well-bound -- such as the present case where $S_p = 14.46$ MeV. We note also that a preliminary cross section was reported for proton removal on a carbon target at 218~MeV/nucleon \cite{Yoshitome21}. Analysis for that system yielded a ground-state to ground-state spectroscopic factor of 0.53(6)~\cite{Tostevin2021-PRC} and so has very minimal overlap with the $(p,2p)$ and earlier NSCL results. A final, published cross-section from this higher energy data set will certainly be of interest in the future.

\section{Experimental Details}

The present experiment was performed at the National Superconducting Cyclotron Laboratory (NSCL) at Michigan State University.  A secondary beam of $^{25}$F was produced following fragmentation of a 140-MeV/nucleon $^{48}$Ca primary beam, accelerated through the Coupled Cyclotron Facility onto a 1034-mg/cm$^{2}$ $^{9}$Be target.  The desired $^{25}$F fragment was separated from other reaction products through the A1900 fragment separator~\cite{A1900}, based on magnetic rigidity and relative energy loss through a 1050~mg/cm$^{2}$ Al wedge.  Fragments were delivered with a momentum acceptance of 1\%~$\Delta p/p$ and impinged upon a 188-mg/cm$^{2}$ thick $^{9}$Be target at the target position of the S800 spectrograph~\cite{S800}.  The residual nuclei were identified event-by-event in the focal plane detectors of the S800, through time-of-flight and energy loss.  The position-sensitive Cathode-Readout Drift Chambers (CRDCs) in the S800 focal plane, when used with an inverse map of the S800 ion-optical elements, enabled a measurement of the parallel momentum distribution of the reaction residues.

While not relevant for the case discussed here, since $^{24}$O has no bound excited states, the target position of the S800 was surrounded by 10 modules of GRETINA~\cite{GRETINA}, populating the most forward positions available with four at $\theta\sim$58$^{\circ}$ and six at $\theta\sim$90$^{\circ}$. The setup enabled exclusive excited-state cross sections to be measured for proton-removal reactions on the $^{21-24}$F isotopes. These data will be discussed in a forthcoming paper.

With a $^{25}$F mid-target energy of $\sim$77~MeV/nucleon, the present proton removal reaction is of sufficiently high beam energy to avoid any significant non-sudden dynamical effects, as have been observed and discussed when well-bound nucleons are removed from a low-energy beam \cite{Flavigny2012}. Such effects might potentially have some limited impact on the earlier, 50 MeV/nucleon removal-reaction experiment of Ref.~\cite{Thoennessen2003}.

\section{Results}

\begin{figure}
\centering
\includegraphics[width=\columnwidth]{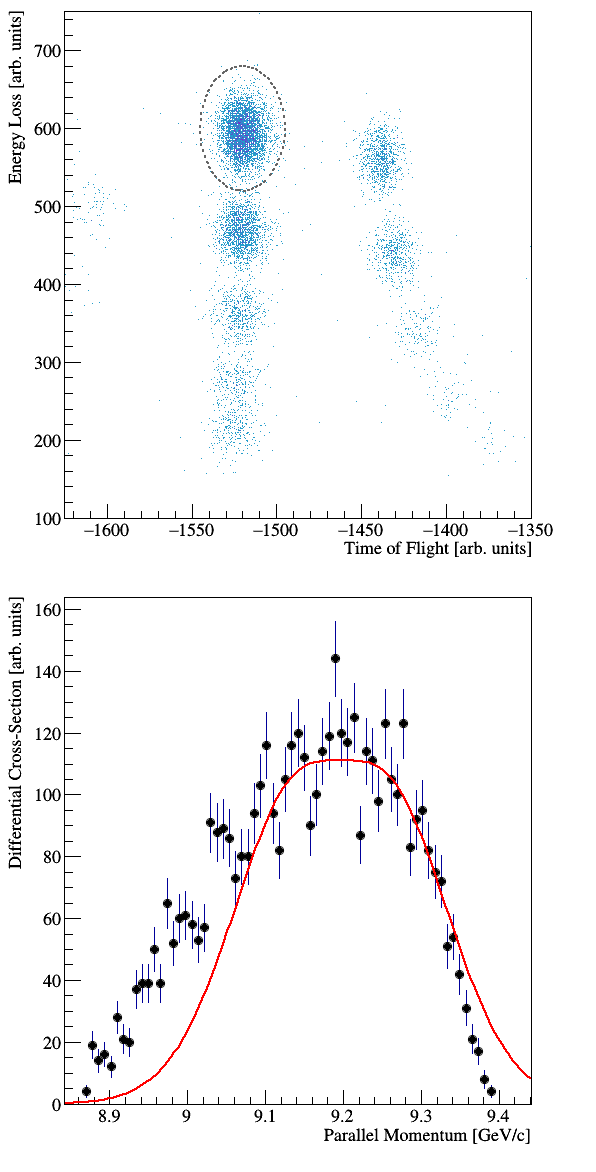}
\caption{(Top Panel) Particle identification plot for reaction residues detected in the S800 focal plane following reaction of incoming $^{25}$F.  The $^{24}$O reaction residues are highlighted in the dashed oval. (Bottom Panel) Parallel momentum distribution for $^{24}$O reaction residues following proton removal from $^{25}$F.  The red line is a calculation of the expected distribution for removal of an $\ell$=2 proton, folded with the experiment response, and shows good agreement with the width of the distribution.  See text for further details.}
\label{fig:experiment}
\end{figure}

A total of 4790(69) $^{24}$O ions were identified in the focal plane of the S800 following the reaction of the incoming $^{25}$F beam, as shown in the upper panel of Figure~\ref{fig:experiment}. Having no particle-bound excited states, all of the observed $^{24}$O nuclei represent proton removals that directly populate the $^{24}$O ground state.  The single-particle cross section, $\sigma_{sp}$, computed with unit spectroscopic factor, and the longitudinal momentum distribution for the $d_{5/2}$ proton removal are calculated using the eikonal-model methodology detailed in Ref. \cite{Tostevin2021-PRC}. The calculated model single-particle cross section is $\sigma_{sp} = 15.7$~mb.

The measured and calculated longitudinal momentum distribution of the $^{24}$O residues is presented in the lower panel of Fig.~\ref{fig:experiment}. The width of the measured distribution is well reproduced by the eikonal-model calculations (solid red line).  As is common, the data display a low momentum tail that arises from more-dissipative reaction events with a greater energy- and momentum-transfer to the target. Such (relatively small) energy transfer is not included in the eikonal-model dynamics. The data also show a cut-off at high momentum -- an instrument acceptance effect in the present experiment. Correcting for this acceptance loss, corresponding to 7(1)\%, as well as the data acquisition livetime, the proton removal cross section from $^{25}$F to the $^{24}$O$_{gs}$ is determined to be $\sigma_{exp}=4.3(6)$~mb.

So, the derived experimental spectroscopic factor of the present work is C$^{2}$S$_{exp}=\sigma_{exp}/\sigma_{sp}= 0.27(4)$, which overlaps the lower part of the error band of the $(p,2p)$ value, 0.36(13)~\cite{Tang2020-PRL}. As stated above, these C$^{2}$S$_{exp}$ are significantly smaller than calculated shell-model spectroscopic factors; for example C$^{2}$S$_{SM}$ = 0.96 for the universal $sd$ (USD) interaction~\cite{Thoennessen2003}. In fact, no state-of-the-art shell-model effective interaction currently predicts such a highly-suppressed $\langle ^{25}$F$|^{24}$O$_{gs},p \rangle$ overlap and C$^{2}$S$_{SM}$ value, all available predictions being close to unity, in line with the independent-particle shell-model expectation. So, the new data presented here, and its unexpected degree of suppression of the nuclear overlap, suggest a significant structural change of the core nucleons of $^{25}$F. In the following discussion we will investigate the potential role of particle-core rotational and/or vibrational coupling degrees of freedom as a driver of such a suppression.

One does not expect that the entire suppression in the proton removal reaction, quantified above, will be attributable to structural changes arising from particle-core or shell-model degrees of freedom. It is recognized that simplified core-coupling models, and even the best available shell-model calculations with their highly-truncated bases, cannot account for a number of few-body, short- and longer-range correlations in the nuclear many-body wave function. These contributions lie outside of the assumed model spaces and/or beyond computational capabilities. Empirically, the magnitudes of such structure-model contributions, and of any systematic limitations of the reaction-dynamics methodology, are encapsulated in the available nucleon removal reaction systematics \cite{Tostevin2021-PRC}. The most recent data compilation now includes reactions of nuclei with a wide range of energies, masses, $np$-asymmetry, and that are quite different structurally. There, comparisons of the experimentally measured inclusive nucleon-removal cross sections ($\sigma_{exp}$) with eikonal-model plus shell-model cross-section calculations for all bound residue final states ($\sigma_{th}$) show a systematic suppression of $R_S=\sigma_{exp}/\sigma_{th}$ with increasing nucleon separation energy. More specifically, the observed $R_S$ suppression varies approximately linearly with $\Delta S$, the difference in separation energies of the removed nucleon (proton or neutron) and that of the other species of nucleon (neutron or proton). Thus, $\Delta S$ provides a measure of the energy asymmetry of the displaced neutron and proton Fermi surfaces~\cite{Tostevin2014-PRC, Tostevin2021-PRC}. Of course, since, for $^{24}$O, the ground-state is the only bound final state, the computed $R_{S}$ value involves only the ground-state to ground-state overlap and its spectroscopic factor.

For the $\Delta S$ value of the present reaction, 10.17 MeV, the now-extensive removal-reaction systematics indicate that $R_S$ is expected to lie in the range 0.45(10) \cite{Tostevin2021-PRC}, whereas the measured (inclusive) cross section of the present experiment and USD shell-model spectroscopic factor of 0.96 derives an $R_{S}$ value of 0.26. To return an $R_S$ value in the expected range, given the measured cross section and the calculated $\sigma_{sp}$, requires a theoretical spectroscopic factor in the range C$^2$S$_{th}$ = 0.56(15). This is the conclusion based on the present  proton-removal reaction data and the available systematics for other systems.

Such systematics considerations for the $(p,2p)$ knockout reaction are less clear. In the work of Tang {et al.} \cite{Tang2020-PRL}, for the single, $^{25}$F-induced reaction, the calculated cross sections are claimed to be absolute. On the other hand, the quasifree $(p,2p)$ measurements and analyses of Atar {\em et al.} (see Figure 4 of \cite{Atar2018}), across the full range of oxygen isotopes, derived ratios of experimental-to-theoretical cross sections with reductions $R$ (analogous to $R_S$) of 0.65(5). We note that the $R_S$ values for the proton-removal reaction data from $^{14}$O and $^{16}$O, and also the $^{16}$O$(e, e'p)$, electron-induced knockout data point (see Fig.1 of \cite{Tostevin2021-PRC}), cases measured using both the $(p,2p)$ and removal-reaction mechanisms, are also consistent with a suppression of order 0.65(5). The situation regarding such systematic effects in the quasifree knockout reaction analyses is thus unresolved at present.

\section{Discussion}

In our earlier works~\cite{AOM2017-PLB,AOM2020-PRC}, we interpreted the structure of both $^{25,29}$F in the framework of the Nilsson plus Particle Rotor model (PRM)~\cite{Nilsson1955,ShapesAndShells,Larsson1978-NPA} where the coupling of a proton $d_{5/2}$ Nilsson multiplet to an effective oxygen core of modest deformation, $\beta_{2}\sim0.16$, can be understood in the rotation-aligned coupling limit. This gives rise to a decoupled band~\cite{Stephens1973-PLB} in agreement with the observed levels where the $d_{5/2}$ strength is fragmented among the ground and excited states, consistent with the reduced $d_{5/2}$ strength in $^{25}$F ground state, observed in the current work.

Given the rather small deformation determined from our previous analysis, it is interesting to apply the Particle Vibration Coupling (PVC) scheme, as developed and discussed by Bohr and Mottelson~\cite{BMVolume2}.  The structure of $^{25}$F could then be considered as a $d_{5/2}$ proton coupled to a quadrupole ($\lambda$ = 2) phonon of frequency $\hbar\omega_2$ in the effective $^{24}$O* core. This approximation is justified by the fact that the proton single-particle levels in a Wood-Saxon potential in this region are characterized by gaps between ($s_{1/2}$, $d_{3/2}$) and the $d_{5/2}$ levels of around $4.3$ and $7$ MeV, respectively~\cite{Volya}, allowing these additional couplings to be ignored.

Following Ref.~\cite{BMVolume2}, the splitting of the quadrupole phonon multiplet is given by:
\begin{equation}
\Delta E(j, I)= \frac{h^2(j,j,2)}{\hbar\omega_2} \Big( \delta_{j I}+ (2j+1)\Big) \Gj{2}{j}{j}{2}{j}{I}
\label{eq:eq1}
\end{equation}
\noindent
and
\begin{equation}
h(j,j,2)=\Big( \frac{5}{4\pi}\Big)^{1/2} \langle j \frac{1}{2} 2 0 | j \frac{1}{2} \rangle \Big( \frac{\hbar\omega_2}{2C_2}\Big)^{1/2} \langle j|\kappa_2(r)|j \rangle
\label{eq:eq2}
\end{equation}
\noindent
In the equations above, $\delta_{j I}$ is the Kronecker delta, \big\{ \big\} is a six-$j$ coefficient, $\langle ~\rangle$ a Clebsch Gordan coefficient, and $\kappa_2(r)=R_0 \partial V/\partial r$ the single-particle radial form factor, giving $\langle j|\kappa_2(r)|j \rangle \approx 50$ MeV. The restoring force parameter, $C_2$, and the phonon frequency, $\hbar\omega_2$, are related to the $E2$ transition probability:
\begin{equation}
B(E2,n=0 \rightarrow n=1) = 5 \Big( \frac{3}{4\pi} ZeR^2 \Big)^{2} \Big(\frac{\hbar\omega_2}{2C_2}\Big)
\label{eq:eq3}
\end{equation}
\noindent

\begin{figure}
\centering
\includegraphics[trim=40 60 60 90, clip,width=0.9\columnwidth, angle=90]{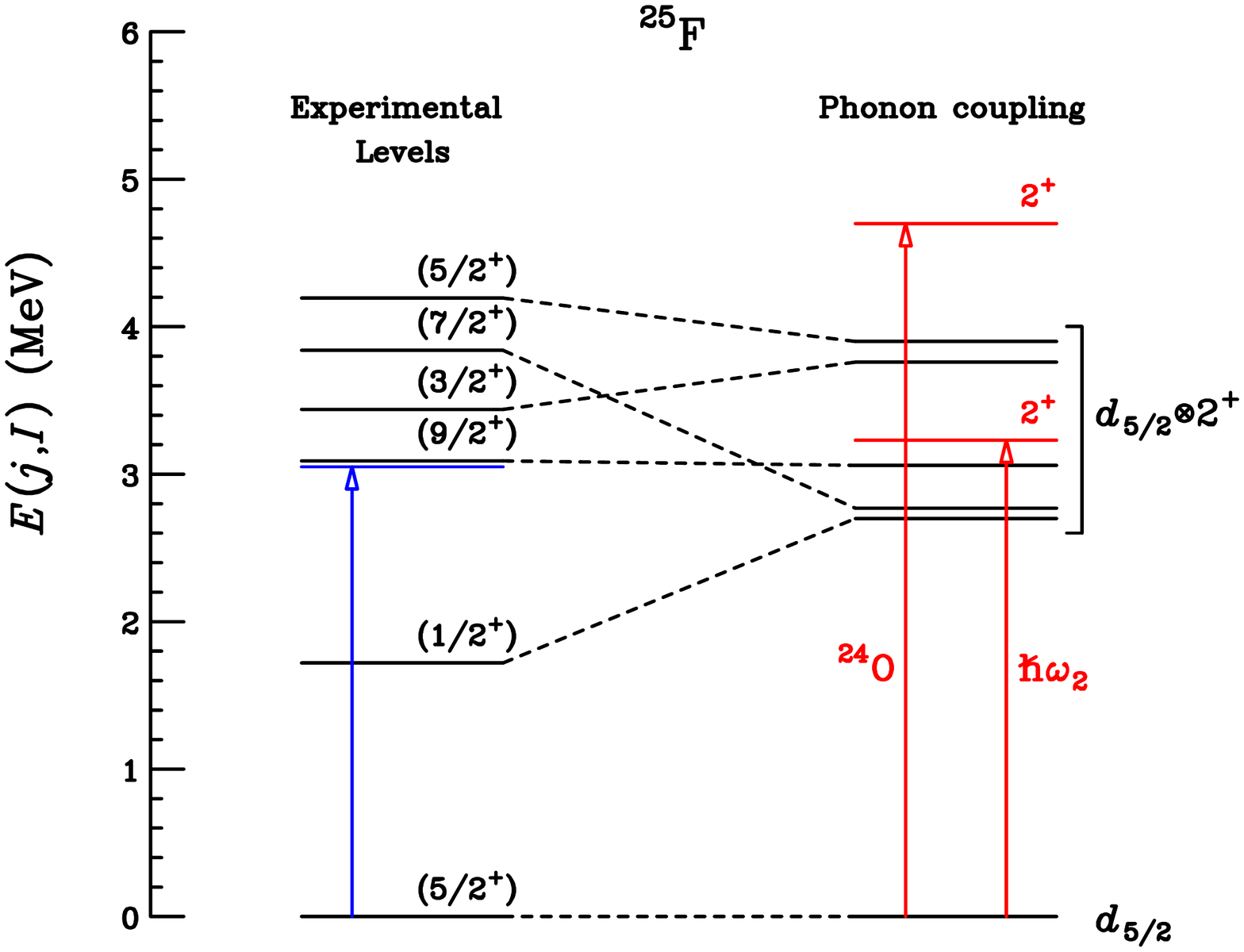}
\caption{ The experimental level scheme of $^{25}$F from Ref.~\cite{Vajta2014-PRC} and the  results of the PVC model showing the phonon splitting at first order.  For reference, the energies of the $2^+$ states in $^{24}$O and of the effective $^{24}$O* core are shown in red. The phonon frequency determined from the spin weighted average of the multiplet  is indicated in blue.}
\label{fig1}
\end{figure}

From a fit to the lowest-energy experimental levels of a given spin $I$ in $^{25}$F~\cite{Vajta2014-PRC}, that we associate with the $\frac{1}{2} \leq I \leq \frac{9}{2}$ multiplet shown in Fig.~\ref{fig1}, we obtain the relevant parameters listed in Table~\ref{table1}, which can be compared with those of the free $^{24}$O nucleus\footnote[2]{The phonon frequency can also be determined from:
%\begin{equation}
$\hbar\omega_2 = \sum_{I=1/2}^{9/2} (2I+1) E(j, I)/ \sum_{I=1/2}^{9/2} (2I+1)$
%\label{eq:eq6}
%\end{equation}
\noindent
which gives 3.05 MeV, consistent with the fitted value of 3.2 MeV.}.

%\noindent
It is seen that the addition of an extra proton tends to soften $^{24}$O, with the effective core becoming more collective, as indicated by the lower phonon energy, stronger coupling term, and smaller restoring force (see Table~\ref{table1}).
A PVC analysis in $^{29}$F, albeit with much less experimental information available, gives the results included in Table ~\ref{table1}, also suggesting a similar behavior of the effective $^{28}$O* core in $^{29}$F.\\

\begin{table}
\centering
\caption{Fitted particle-vibration coupling parameters for the effective $^{24}$O* core in $^{25}$F, compared to those in $^{24}$O.  Similarly for the $^{28}$O-$^{29}$F case.  Phonon energies, coupling constants and restoring force parameters are in MeV and $B(E2)$'s in $e^2b^2$.}
\bigskip
\begin{tabular}{c|c|c|c}
\hline\hline
 Parameter   &  $^{24}$O & $^{24}$O*& $^{28}$O*  \\
\hline
$\hbar\omega_2  $ & 4.7 & 3.2 & 2.0 \\
$h(j,j,2)$ &0.73&1.58 &1.75 \\
$C_2$  &204 &140  &97 \\
\hline
$B(E2)$  &0.0012&0.0055&0.0082\\
\hline\hline
\end{tabular}
\label{table1}
\end{table}

Within the PVC scenario, the single-particle state is renormalized to:
\begin{equation}
\widehat{| j \rangle} \approx  a | j,n=0 \rangle + b|j,n=1 \rangle
\label{eq:eq4}
\end{equation}
\noindent
with
\begin{equation}
b= \frac{h(j,j,2)}{\hbar\omega_2 }
\label{eq:eq5}
\end{equation}
\noindent
as a result of its coupling to the phonon.  

Thus the spectroscopic factor for one-proton removal from the $5/2^+$ ground state of $^{25}$F to the ground state of $^{24}$O is:
\begin{equation}
C^2S_{PVC}= a^2\langle ^{24}\text{O} | ^{24}\text{O*}\rangle^2
\label{eq:eq6}
\end{equation}
\noindent
where, from Eq. (\ref{eq:eq4}), $a^2$ accounts for the $\widehat{| d_{5/2} \rangle}$ single-particle renormalization.  We estimate the overlap between the initial and final cores, $\langle ^{24}\text{O} | ^{24}\text{O*}\rangle \approx 0.87$,  by the overlap of harmonic oscillator wavefunctions in $\beta_2$, each adjusted to give the root-mean-squared deformation $\sqrt{\langle \beta_2^2 \rangle}$ obtained from the corresponding $B(E2)$ values in Table~\ref{table1}. With the above, we obtain C$^2$S$_{th, PVC}$= 0.6, consistent with the reduced overlap between the $^{24}$O$^{*}$ effective core in $^{25}$F and the bare $^{24}$O nucleus that has been inferred from the data.  Within the PVC calculation, the remaining part of the $d_{5/2}$ strength, $b^2$, gets distributed among the first and other excited state resonances in $^{24}$O that decay to $^{23}$O.

\section{Conclusion}
We have reported a new measurement of the proton removal reaction from $^{25}$F on a $^9$Be target at 77 MeV/nucleon, performed at the NSCL using the S800 spectrometer. The measured and calculated cross sections for the $^{25}$F$(5/2^+$) ground-state to $^{24}$O$(0^+$) ground-state proton removals reveal a substantial depletion of the proton $d_{5/2}$ strength, with an experimental spectroscopic factor $C^2S_{exp}= 0.27(1)$. A similar depletion was reported in Ref.~\cite{Tang2020-PRL} from an inverse-kinematics, quasifree $(p,2p)$ knockout-reaction measurement. Our new result agrees with that from an updated analysis of the lower-energy removal-reaction measurement of Ref.~\cite{Thoennessen2003}, made on a carbon target. Collectively, these data sets indicate a significant reduction of the $\langle ^{25}$F$|^{24}$O$_{gs},p \rangle$ overlap compared to shell-model expectations and hence that the core nucleons of $^{25}$F do not behave as the free, more-rigid, doubly-magic $^{24}$O nucleus.

Taking into account the well-documented systematic dependence of the ratios of the measured and calculated inclusive nucleon-removal reaction cross sections, $R_S$, upon the difference between the separation energies of the two nucleon species from the projectile, $\Delta S$, the newly presented data are consistent with a theoretical spectroscopic factor of C$^2$S$_{th}$ = 0.56(15).

Expanding upon the Particle Rotor model (PRM) interpretation of the $^{25}$F structure of Ref.~\cite{AOM2020-PRC}, here we have considered the coupling of a $d_{5/2}$ proton to a quadrupole vibrational core. Unsurprisingly, the Particle Vibration Coupling (PVC) approach also provides a good description of the experimental data by requiring an effective $^{24}$O* core with a phonon energy, $\hbar\omega_2$= 3.2 MeV, and a $B(E2) \approx 2.7$ W.u. -- softer and more collective than the bare $^{24}$O nucleus. Due to the $\widehat{| d_{5/2} \rangle}$ single-particle renormalization, and the reduced overlap between the free and effective cores, we obtain a reduced spectroscopic factor C$^2$S$_{th, PVC}$= 0.6 in agreement with the C$^2$S$_{th}$ value required by the data. Both the Nilsson deformed mean field and the PVC approaches appear to capture the essential properties of the effective core in $^{25}$F, suggesting that the additional $d_{5/2}$ proton tends to polarize the free, doubly-magic $^{24}$O in such a way that it becomes either slightly deformed (PRM) or a quadrupole vibrator (PVC).  A similar behavior appears to be at play in the $^{28}$O-$^{29}$F system, although experimental data is needed to confirm that case.

 \begin{acknowledgments}
We would like to thank the operations team at NSCL for their work in beam delivery during the experiment.  AOM would like to thank Roberto Liotta for discussions on particle-vibration-coupling.  GRETINA was funded by the U.S. DOE Office of Science.  Operation of the array at NSCL was supported by DOE under Grants No. DE-SC0014537 (NSCL) and DE-AC02-05CH11231 (LBNL).  This material is based upon work supported by the U.S. Department of Energy, Office of Science, Office of Nuclear Physics under Contract  Nos. DE-AC02-05CH11231 (LBNL), DE-AC05-00OR22725 (ORNL) and Grant Nos. DE-SC0020451 (NSCL), DE-FG02-97ER41041 (UNC) and DE-FG02-97ER41033 (TUNL).  Support also came from the U.S. National Science Foundation (NSF) under Cooperative Agreement No. PHY-1565546 (NSCL).  This work was also supported by the UK STFC under contract Nos. ST/V001108/1, ST/P003885/1 and ST/L005727/1.  MP acknowledges support by the Royal Society.
%\clearpage
\bigskip
\bigskip

\end{acknowledgments}

\bibliography{references}

\end{document}